\documentclass{pos}

\usepackage{amssymb,amsmath}

\usepackage{slashed}
\usepackage{graphicx}

\newcommand{\nn}{\nonumber \\ }
\def\rgel{\mathsf{L}}
\def\bn{ \bar n}

\newcommand{\beq}{\begin{equation}}
\newcommand{\eeq}{\end{equation}}
\newcommand{\bea}{\begin{eqnarray}}
\newcommand{\eea}{\end{eqnarray}}
\newcommand{\co}{\, ,}
\newcommand{\fs}{\, .}

\newcommand\Lq{\mathsf{L_Q}}
\newcommand\Lm{\mathsf{L_M}}

\title{Using SCET to calculate electroweak corrections in gauge boson production}

\ShortTitle{Using SCET to calculate electroweak corrections in gauge
  boson production} 
\author{Jui-yu Chiu\\
Department of Physics, University of California at San Diego,
  La Jolla, CA 92093\\
  E-mail: \email{jchiu@physics.ucsd.edu}}

\author{\speaker{Andreas Fuhrer}\\
Department of Physics, University of California at San Diego,
  La Jolla, CA 92093\\
  E-mail: \email{afuhrer@physics.ucsd.edu}}

\author{Andr\'e~H.~Hoang\\
Max-Planck-Institut f\"ur Physik (Werner-Heisenberg-Institut), 
F\"ohringer Ring 6,\\ 80805 M\"unchen, Germany\\
  E-mail: \email{ahoang@mppmu.mpg.de}
}

\author{Randall Kelley\\
Department of Physics, University of California at San Diego,
  La Jolla, CA 92093\\
  E-mail: \email{rkelley@physics.ucsd.edu}}

\author{Aneesh V.~Manohar\\
Department of Physics, University of California at San Diego,
  La Jolla, CA 92093\\
  E-mail: \email{amanohar@ucsd.edu}}

\abstract{We extend an effective theory framework developped in
  Refs.~\cite{CGKM,CKM} to sum
  electroweak Sudakov logarithms in high energy processes to also
  include massive gauge bosons in the final state. 
 The calculations require an additional regulator on top of
 dimensional regularization  to tame the collinear
  singularities. We propose to use the $\Delta$ regulator, which
  respects soft-collinear factorization.}

\FullConference{To be published in the proceedings of: \\
International Workshop on Effective Field Theories: from the pion to the upsilon \\
		February 2-6 2009\\
		Valencia, Spain}

\begin{document}

\section{Introduction}

The Large Hadron Collider (LHC) has a center-of-mass energy of $\sqrt
s =14$~TeV, and will be able to measure collisions with a partonic
center-of-mass energy of several TeV, more than an order of magnitude
larger than the masses of the electroweak gauge bosons. Radiative
corrections to scattering processes depend on the ratio of mass
scales, and radiative corrections at high energy depend on logarithms
of the form $\log s/M^2_{W,Z}$.  In high energy exclusive processes,
radiative corrections are enhanced by two powers of a large logarithm
for each order in perturbation theory, and the logarithms are often
referred to as Sudakov (double) logarithms.  Electroweak Sudakov
corrections are not small at LHC energies, since $\alpha \log^2
s/M^2_{W,Z}/(4 \pi \sin^2 \theta_W) \sim 0.15$ at
$\sqrt{s}=4$~TeV. These Sudakov corrections lead to sizeable effects and might be summed to all orders. 

The Sudakov logarithm $\log(s/M_{W,Z}^2)$ can be thought of as an
infrared logarithm in the electroweak theory, since it diverges as
$M_{W,Z}\to0$. By using an effective field theory (EFT), these
infrared logarithms in the original theory can be converted to
ultraviolet logarithms in the effective theory, and summed using
standard renormalization group techniques. The effective theory needed
is soft-collinear effective theory (SCET)~\cite{BFL,SCET}, which has
been used to study high energy processes in QCD, and to perform
Sudakov resummations arising from radiative gluon corrections. 

The summation of electroweak Sudakov logarithms using effective field
theory methods has extensively been discussed in Ref.~\cite{CGKM} for the Sudakov form
factor and in Ref.~\cite{CKM} for four fermi scattering processes. Here, we
extend the discussion to processes with gauge bosons in the final
state. Only the Sudakov form factor will be
considered in the following.

\section{Exponentiation}
We start by summarizing some known properties of the Sudakov
form-factor~\cite{collins} for the vector current. The Euclidean form-factor $F_E(Q^2)$ has the
expansion ($\rgel=\log(Q^2/M^2)$) 
\begin{eqnarray}
F_E &=& 1 + \alpha\left( k_{12}\rgel^2+k_{11}\rgel + k_{10}\right)
+\alpha^2\left(k_{24}\rgel^4+k_{23}
\rgel^3+k_{22}\rgel^2+k_{21}\rgel+k_{20}\right) + \ldots \, ,
\label{1}
\end{eqnarray}
with the $\alpha^n$ term having powers of $\rgel$ up to
$\rgel^{2n}$. In the literature, the highest power of $\rgel$ is
called the $\text{LL}_{\text{F}}$ term, the next power is called the
$\text{NLL}_{\text{F}}$ term, etc.\ We have included the subscript $F$
(for the form-factor) to distinguish it from the renormalization group
counting described below. 

The series for $\log F_E(Q^2)$ takes a simpler form
\begin{eqnarray}
\log F_E &=&  \alpha\left(\tilde k_{12}\rgel^2+\tilde k_{11}\rgel +
\tilde k_{10}\right) +\alpha^2\left(\tilde k_{23}
\rgel^3+ \tilde k_{22}\rgel^2+ \tilde k_{21}\rgel+\tilde
k_{20}\right) +\ldots \co
\label{2}
\end{eqnarray}
with the $\alpha^n$ term having powers of $\rgel$ up to $\rgel^{n+1}$, and the expansion begins at order $\alpha$. Note that Eq.~(\ref{2}) implies non-trivial relations among the coefficients $k_{nm}$ in Eq.~(\ref{1}). At order $n$, there are $2n+1$ coefficients $k_{nm}$, $0 \le m \le 2n$ in Eq.~(\ref{1}), but only $n+2$ coefficients $\tilde k_{nm}$, $0 \le m \le n+1$ in Eq.~(\ref{2}).

The right-hand-side of Eq.~(\ref{2}) can be written in terms of the LL series $\rgel f_0(\alpha \rgel)=\tilde k_{12} \alpha \rgel^2+
\tilde k_{23} \alpha^2 \rgel^3 + \ldots$, the NLL series $f_1(\alpha \rgel)=\tilde k_{11} \alpha \rgel +
\tilde k_{22} \alpha^2 \rgel^2 + \ldots$, the NNLL series $\alpha f_2(\alpha \rgel)=\tilde k_{10} \alpha  +
\tilde k_{21} \alpha^2 \rgel + \ldots$ etc.\ as
\begin{eqnarray}
\log F_E &=&\rgel f_{0}(\alpha \rgel)+f_{1}(\alpha \rgel)+\alpha f_{2}(\alpha \rgel)
+ \ldots\, .
\label{3}
\end{eqnarray}
$f_0$ and $f_1$ begin at order $\alpha$, and the remaining $f_n$ begin at order one.

Here, LL, NLL, etc.\ (with no subscripts) will refer to the
counting for $\log F_E$. This is also the counting appropriate for a
renormalization group improved computation, and is different from the
conventional counting discussed above. If one looks at the order
$\alpha^2$ terms, for example, the conventional counting is that the
$\rgel^4$ term is $\text{LL}_{\text{F}}$, the $\rgel^3$ term is
$\text{NLL}_{\text{F}}$, the $\rgel^2$ term is
$\text{N}^2\text{LL}_{\text{F}}$, the $\rgel$ term is
$\text{N}^3\text{LL}_{\text{F}}$, and the $\rgel^0$ term is
$\text{N}^4\text{LL}_{\text{F}}$. Using our counting, the terms are
given by exponentiating $\log F_E$ to $\text{LL}$,  $\text{NLL}$,
$\text{N}^2\text{LL}$, $\text{N}^2\text{LL}$, and
$\text{N}^3\text{LL}$, respectively. At higher orders, the mismatch in
powers of N between the two counting methods increases.
 
Since for electroweak corrections at the TeV scale $\alpha\rgel^2$ is
quite sizeable, the $\text{LL}$ series might be summed up to all orders.

\section{SCET formalism}\label{sec:scet}

The theory we consider is a $SU(2)$ spontaneously broken gauge theory, with a
Higgs in the fundamental representation, where all gauge bosons have a common
mass, $M$. It is convenient to write the group
theory factors using $C_F$, $C_A$, $T_F$\footnote{Note that the results only
  hold for $C_A=2$, since for an $SU(N)$ group with $N>2$, a fundamental Higgs
  does not break the gauge symmetry completely.}  

The physical quantity of interest is the Sudakov form factor $F_O(Q^2)$ in the
Euclidean region,  
\beq\label{eq:sff}
F_O(Q^2) = \mathcal{N} \langle p_2 | O | p_1 \rangle \co
\eeq
where $Q^2 = -(p_2-p_1)^2 \gg M^2$, $O$ is a generic operator and
$\mathcal{N}$ a normalization factor.
In SCET, $F_O(Q^2)$ is computed in three steps: (i) matching from the
full gauge theory to SCET at $\mu=Q$ (high-scale matching) (ii) running in
SCET between $Q$ and $M$ and (iii) integrating out the gauge bosons at $\mu=M$
(low-scale matching). All computations are done to leading order in
SCET power counting, i.e.\ neglecting $M^2/Q^2$ power corrections. 

The SCET fields and Lagrangian depend on two null four-vectors $n$ and $\bar
n$,  with $n=(1,\bf{n})$ and $\bn=(1,-\bf{n})$, where $\bf{n}$ is a unit
vector, so that $\bar n \cdot n=2$. In the Sudakov problem, one works in the
Breit frame, with $n$ chosen to be along the $p_2$ direction, so that $\bar n$
is along the $p_1$ direction. In the Breit frame, the momentum transfer $q$
has no time component, $q^0=0$, so that the particle is back-scattered.
The light-cone components of a four-vector $p$ are defined by $p^+ \equiv n
\cdot p$, $p^- \equiv \bn \cdot p$, and $p_\perp$, which is orthogonal to $n$
and $\bn$, so that 
\begin{eqnarray}
p^\mu &=& \frac 12 n^\mu (\bn \cdot p)+\frac12 \bn^\mu (n \cdot p) + p_\perp^\mu.
\label{4}
\end{eqnarray}
In our problem, $p_1^-=p_{1\perp}=p_2^+=p_{2\perp}=0$, and $Q^2=p_1^+ p_2^-$.
A gauge boson moving in a direction close to $n$ is described by the $n$-collinear
SCET field $A_{n,p}(x)$, where $p$ is a label momentum, and has components
$\bar n \cdot p$ and $p_\perp$~\cite{BFL,SCET}.  It describes particles (on-
or off-shell) with energy $2E\sim\bar n \cdot p$, and $p^2 \ll Q^2$. The total momentum of the field $A_{n,p}(x)$ is $p+k$, where $k$ is the residual momentum of order $Q \lambda^2$ contained in the Fourier
transform of $x$. The scaling of the momenta is $\bar n \cdot p \sim Q$, $n \cdot p \sim Q
\lambda^2$, $p_\perp \sim Q \lambda$. The $\bar n$-collinear field
$A_{\bar n,p}(x)$ contains massive gauge bosons 
moving near the $\bar n$-direction, with momentum scaling $n \cdot p \sim Q$,
$\bar n \cdot p \sim Q \lambda^2$, $p_\perp \sim Q \lambda$. Here we have
$\lambda\sim M/Q$,  where $\lambda \ll 1$ is the power counting parameter used for the
EFT expansion. The mass-mode 
field (see Ref.~\cite{fhms}) contains massive gauge bosons with all momentum components scaling as 
$Q \lambda\sim M$.

\section{Wilson lines and regulator}\label{sec:wr}

Before we outline how to extend the framework of Refs.~\cite{CGKM,CKM} to also
include final state gauge bosons, we elaborate on a technical issue
encountered when calculating SCET diagrams with a massive gauge
boson. 

Consider a high energy scattering process with two or more particles, in the
$n_i$ direction, $i=1,\ldots ,r$. 
$n_i$-collinear gauge bosons, which have momentum parallel to particle~$i$ can
interact with particle~$i$, or with the other particles $j\not=i$. The
coupling of $n_i$-collinear gauge bosons to particle~$i$ is included
explicitly in the SCET Lagrangian. The particle-gauge interactions are
identical to those in the full theory, and there is no simplification on
making the transition to SCET. However, if an $n_i$-collinear gauge boson
interacts with a particle $j$ not in the $n_i$-direction, then
particle~$j$ becomes off-shell by an amount of order $Q$, and the intermediate
particle~$j$ propagators can be integrated out, giving a Wilson line
interaction in SCET. The form of these operators was derived in
Ref.~\cite{SCET,SCET2}, and gives the Wilson line interaction $W_{n_i}^\dagger
\xi_{n_i}$, where $W_{n_i}$ is a Wilson line in the $\bn_i$ direction in the
same representation as $\xi_{n_i}$. This is easy to see in processes with only
two collinear particles. But even in complicated scattering processes with
more than two collinear particles the Wilson line interaction still has
the form $W_{n_i}^\dagger \xi_{n_i}$.
To see that this statement also holds at one loop is not
straightforward. The reason is that loop diagrams require an
additional regulator on top of dimensional regularization. This
introduces a dependence on all the other collinear directions $n_j$. Here, we
use the $\Delta$ regulator introduced in Ref. \cite{CFHKM}, which amounts
to modify the propagator denominators as
\begin{eqnarray}
\frac{1}{(p_i+k)^2-m_i^2} \to \frac{1}{(p_i+k)^2-m_i^2-\Delta_i}.
\label{deltareg}
\end{eqnarray}
In SCET, the collinear propagator denominators have the replacement of
Eq.~(\ref{deltareg}). Accordingly, Wilson lines become
\begin{eqnarray}
\frac{\epsilon \cdot n_j}{k \cdot n_j} &\to &
\frac{\epsilon \cdot \bar n_i}{k \cdot \bar n_i - \delta_{j,n_i}}\,,\nn
\delta_{j,n_i} &\equiv& \frac{2\Delta_j}{(n_i \cdot n_j)(\bn_j \cdot p_j) }.
\label{eq7}
\end{eqnarray} 
It turns out (see Ref.~\cite{CFHKM}) that after zero-bin subtraction
\cite{MS} (see also Ref.~\cite{CH}), the
dependence on the other collinear directions drop out and
$n_i$-collinear gauge boson emission can be combined into a single
Wilson line. Note that in an intermediate step, regulator dependent
regions are introduced into the calculation.  However, they cancel
between the soft- and collinear diagrams. The sum of all diagrams is
of course independent of the $\Delta$ regulator.

\section{Transversely polarized $W$ bosons}

Consider scattering of two gauge bosons via the operator $O_T
= F_{\mu\nu}^a F^{\mu \nu,a}$ with $F_{\mu \nu} = \partial_\mu A_\nu^a -
\partial_\nu A_\mu^a + gf^{abc}A_\mu^b A_\nu^c$ the non-abelian field
strength tensor,
\begin{equation}\label{eq:norm}
\left<p_2|O_T|p_1\right> = 4 F_{O_T}(Q^2) \left[p_1\cdot p_2
  \,\epsilon(p_1)\cdot\epsilon^*(p_2)- \epsilon^*(p_2)\cdot p_1 \,
  \epsilon(p_1)\cdot p_2 \right] .
\end{equation}
Following the steps outlined in Sect.~\ref{sec:scet}, one first
matches the full theory matrix element at the high scale $\mu \sim Q$ onto
the matrix element in the effective theory. The corresponding
operator to $O_T$ in the effective theory reads $\tilde{O}_T =
B_{n,p_2}^{\dagger \perp} B_{\bar{n},p_1}^\perp $, with $B^\mu_{n,p}$ defined as \cite{AKS,BCO}
\begin{align}
B^\mu_{n,p} &= \frac{1}{g}\left[W_n^\dagger iD_n^\mu W_n \right],
&iD_n^\mu &= i\partial^\mu-gA_{n,p}^\mu \fs
\end{align}
The matching coefficient $C_T(\mu)$ of the effective theory operator up
to $\mathcal{O}(\alpha)$ reads
\beq
C_T(\mu) = 2Q^2 \left[ 1+\frac{\alpha}{4\pi} C_A \left(
  - \Lq^2 +\frac{\pi^2}{6} \right)
  \right] \co
\eeq
\begin{figure}
\centering
\begin{tabular}{ccc}
\includegraphics[height=3.5cm]{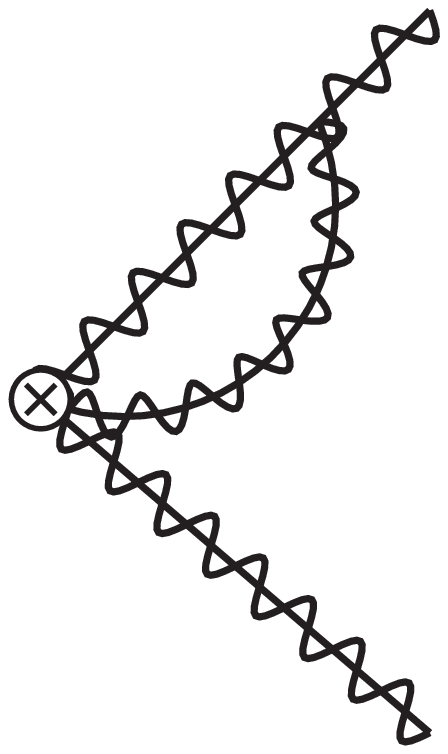}&\includegraphics[height=3.5cm]{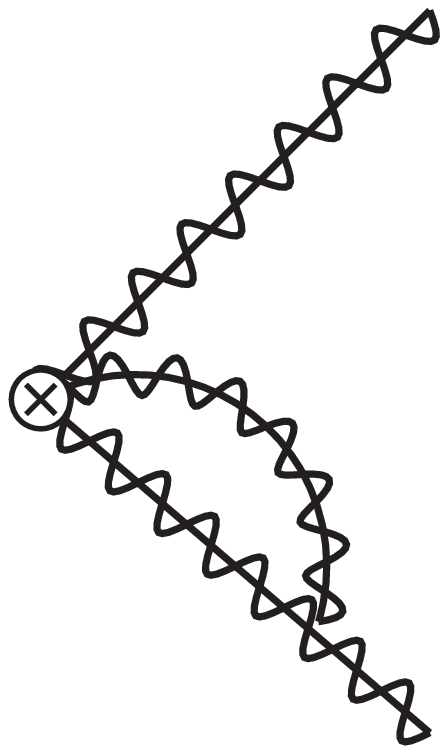}&\includegraphics[height=3.5cm]{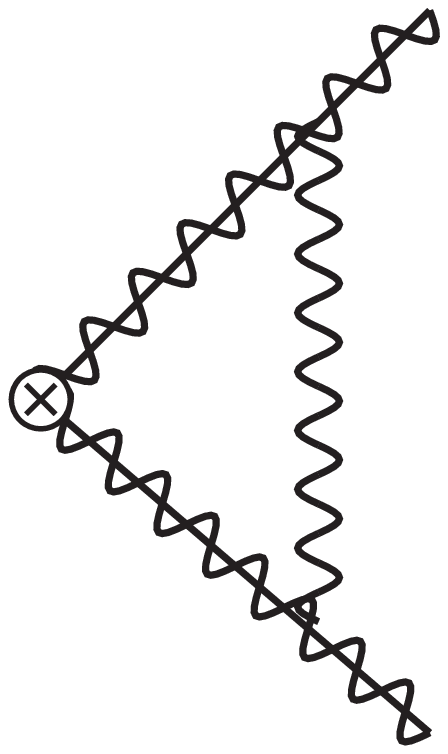}
\end{tabular}
\caption{One loop diagrams in the effective theory. The wiggly line
  denotes a mass mode gauge boson and a wiggly line with an overlaid straight line
  indicates a collinear field. Wave function diagrams are not shown. \label{fig:1loop}}
\end{figure}
where we use the notation $\mathsf{L_X} \equiv \ln\left(X^2/\mu^2 \right)$. This is just the finite part of the one loop amplitude in
the full theory, normalized according to Eq.~(\ref{eq:norm}). Note that since we
are in the high energy limit, the $SU(2)$ gauge theory is in the
unbroken phase and all the masses have been set to zero. The effective theory one loop diagrams
shown in Fig.~\ref{fig:1loop} are all scaleless in dimensional regularization and vanish.
The anomalous dimension of the effective theory operator $\tilde{O}_T$
follows from the $\frac{1}{\epsilon}$ poles from the one loop matrix
element,
\beq
\gamma_T = \frac{\alpha}{4\pi}C_A \left[ 4 \Lq -4 \right]+ \frac{\alpha}{2\pi}\left[-\frac{10}{3}C_A+\frac{4}{3}T_Fn_f+\frac{2}{3}T_Fn_s+\frac{1}{3}
  \right]    \fs
\eeq
The second term in the square brackets is the contribution from the wave function
renormalization of the gauge bosons.
The matching coefficient $C_T(\mu)$ can be evolved down to a scale $\mu
\sim M$ with the renormalization group equation
\beq
C_T(\mu_2) = C_T(\mu_1) \exp\left[\int_{\mu_1}^{\mu_2}
  \frac{d\mu}{\mu}\gamma_T(\mu) \right] \fs
\eeq

At the low scale $\mu \sim M$, the gauge bosons are integrated out by
matching the effective theory with dynamical massive gauge bosons onto
an effective theory where the gauge bosons are treated as heavy
background fields. The matching coefficient at the low scale,
$D_T(\mu)$, up to order $\mathcal{O}(\alpha)$ is again obtained by comparing the
finite parts of the one loop matrix elements. Since the gauge boson
has to be treated as massive, the one loop diagrams do not vanish
anymore. The calculation of the individual diagrams shown in
Fig.~\ref{fig:1loop} requires an additional regulator, as described in
Sect.~\ref{sec:wr}. In the effective theory below the scale $M$, there
are no dynamical interacting degrees of freedom left and therefore, no
quantum corrections appear. Also, there is no need to evolve $D_T(\mu)$.
One obtains
\bea
D_T(\mu) &=& 1+\frac{\alpha}{4\pi} C_A \left[2 \Lm \Lq-\Lm^2-2\Lm+2
  -\frac{5\pi^2}{6}+2f_s(1,1) \right]+\delta R_W \co \nn
f_s(v,w) &=& \int_0^1 dx \frac{2-x}{x}\ln\frac{1-x+wx-vx(1-x)}{1-x} \fs
\eea
The quantity $\delta R_\chi$ denotes the finite part of the residue of
the full propagator of the field $\chi$. 
The result for the resummed Sudakov form factor at high energies reads
\beq
F_{O_T}(Q^2) = C_T(Q) \exp \left[\int_{Q}^{M}
  \frac{d\mu}{\mu}\gamma_T(\mu) \right] D_T(M) \fs
\eeq
Note that it was proven in Ref.~\cite{CGKM} that there can appear at most one
logarithm of the high scale, $\Lq$, in the low scale matching
coefficient $D_T$. 

\section{Longitudinally polarized $W$ bosons}
The emission of longitudinally polarized gauge bosons at high energies
is related to a truncated matrix element (indicated by the subscript '$\mathrm{tr}$') of unphysical Goldstone bosons by
virtue of the Goldstone boson equivalence theorem (see Ref.~\cite{ET}),
\beq
\epsilon^\mu(p_1) \epsilon^{\nu *}(p_2) R_W \big<0|T A^a_\mu A^b_\nu
|0\big>_{\mathrm{tr}} = i^2\mathcal{E}^2 R_\phi\big<0|T\phi^a
\phi^b|0\big>_{\mathrm{tr}}+\mathcal{O}\left(\frac{M}{E} \right) \fs
\eeq
The quantity $\mathcal{E}$ is a nontrivial correction factor arising
from the truncation of the Greens functions,
\beq\label{eq:E}
\mathcal{E} =
1+\frac{\alpha}{4\pi}\Bigg[-\frac{z^2}{4}+\frac{47\pi}{12
    \sqrt{3}}-\frac{73}{12}+\frac{2z^4-7z^2+5
  }{8}\ln(z)+\frac{-2z^5+7z^3+z}{4\sqrt{4-z^2}}
  \arctan\left(\frac{\sqrt{2-z}}{\sqrt{2+z}} \right) \Bigg]
\eeq
with $z = M_h/M$ the Higgs-Goldstone boson mass ratio. Note that
$\mathcal{E}$ does not run, however, it compensates the gauge
dependence of the unphysical Goldstone bosons.

The effective theory calculation proceeds in the same manner as
described in the previous section. The full theory operator is the
square of the Higgs doublet operator
\begin{align}
O_L &= H^\dagger H \co &H &= \frac{1}{\sqrt{2}}\left( \begin{array}{c}
\phi^2+i\phi^1 \\
h-i\phi^3 \\ \end{array} \right) \fs
\end{align}
Since the Higgs doublet field in the effective theory is a scalar
collinear field $\phi_{n,p}$ in the same representation as $H$, the
high scale matching coefficient $C_L(\mu)$ of the operator in the
effective theory, $\tilde{O}_L = [\phi_{n,p_2}^\dagger W_n][W_{\bar{n}}^\dagger
  \phi_{\bar{n},p_1} ]$, and its anomalous dimension $\gamma_L$ have already been
calculated in Ref.~\cite{CGKM}. The result is
\begin{align}
C_L(\mu) &= 1+\frac{\alpha}{4\pi}C_F
\left[-\Lq^2+4\Lq-2+\frac{\pi^2}{6}\right]\co &\gamma_L &=
\frac{\alpha}{4\pi}C_F\left[4\Lq-8 \right]\fs
\end{align}
At the low scale, the $SU(2)$ invariant operator splits up into
invariants of the remaining $SO(3)$ custodial symmetry. Again, in the
theory below $\mu \sim M$, the Higgs and the Goldstone boson are
treated as heavy fields $h_v^{(h)}$ and $h_v^{(\phi)}$ and the theory has no quantum
corrections. Matching onto the operators $O_{hh} = h_{v_2}^{(h)} h_{v_1}^{(h)}$ and
$O_{\phi \phi} = h_{v_2}^{(\phi)}h_{v_1}^{(\phi)}$, one finds
\bea
D_L^{(\phi \phi)}(\mu) &=& 1 + \frac{\alpha}{4\pi}C_F\left[2\Lm
    \Lq-\Lm^2-2\Lm+2-\frac{5\pi^2}{6}+\frac{4}{3}f_s(1,1)+\frac{2}{3}f_s(1,z^2)
    \right] + \delta R_\phi \co \nn
D_L^{(hh)}(\mu) &=& 1 + \frac{\alpha}{4\pi} C_F \left[2\Lm
    \Lq-\Lm^2-2\Lm+2-\frac{5\pi^2}{6}+ 2 f_s(z^2,1)\right] + \delta
R_h \fs
\eea
Having resummed the large logarithms in the Goldstone boson
scattering off the operator $O_L$, one only needs to correct with the factor $\mathcal{E}$ of
Eq.~(\ref{eq:E}) to obtain the result for the scattering of
longitudinally polarized gauge bosons.

\section{Summary and conclusions}
We discuss a new regulator to tame the collinear singularities in the
effective theory with massive gauge bosons. This regulator respects
soft-collinear factorization. 
Furthermore, the framework of Refs.~\cite{CGKM,CKM} is extended
to also include gauge bosons in the final state. We restrict the
discussion to a spontaneously broken $SU(2)$ gauge symmetry. Results for
scattering of gauge bosons off an external operator are presented. We
consider massive gauge bosons with transverse as well as longitudinal polarization.
The latter relies on the Goldstone boson equivalence theorem.
A generalization to the full standard model gauge group is
straightforward.

\end{document}